\documentstyle[12pt,epsf]{article}
\begin{document}
\begin{titlepage}

    \title{Maximal Acceleration Effects in Reissner-Nordstr\"om Space.}

\author{V.Bozza$^{a,b}$\thanks
{E-mail: valboz,scarpetta@physics.unisa.it},
 A. Feoli$^{b,c}$\thanks{E-Mail: Feoli@unisannio.it},
  G. Papini$^{d,e}$\thanks{E-mail:
papini@uregina.ca}, G. Scarpetta$^{a,b,f}$}
\date{\empty}
\maketitle
\centerline{\em $^a$ Dipartimento di Scienze Fisiche ``E.R. Caianiello'', Universit\`a di Salerno, Italy.}

 \centerline{\em $^b$ Istituto Nazionale di Fisica Nucleare, Sezione di Napoli.}

 \centerline{\em $^c$ Facolt\`a d'Ingegneria, Universit\`a del Sannio, Benevento, Italy.}

 \centerline{\em $^d$ Department of Physics, University of Regina, Regina, Sask. S4S 0A2, Canada.}

 \centerline{\em $^e$ Canadian Institute for Theoretical Astrophysics,
 University of Toronto,}
 \centerline{\em Toronto, Ontario K7L 3N6, Canada}
 \centerline{\em $^f$ IIASS, Vietri sul Mare (SA), Italy}

\bigskip
\begin{abstract}

The dynamics of a relativistic particle in a Reissner-Nordstr\"om
background is studied using Caianiello model with maximal
acceleration. The behaviour of the particle, embedded in a new
effective geometry, changes with respect to the classical scenario
because of the formation of repulsive potential barriers near the
horizon. Black hole formation by accretion of massive particles is
not therefore a viable process in the model. At the same time, the
naked singularity remains largely unaffected by maximal
acceleration corrections.

\end{abstract}

\thispagestyle{empty}  PACS: 04.50.+h, 04.70.Bw

     \vfill
     \end{titlepage}

A model developed by Caianiello and collaborators \cite{qg}
provides quantum mechanics with a geometrical framework and delves
into a number of fundamental issues such as the unification of
general relativity and quantum mechanics and the regularization of
field equations. In particular, the model interprets
 quantization as
curvature of the eight-dimensional space-time tangent bundle TM,
incorporates the Born reciprocity principle and the notion that
the proper acceleration of massive particles has an upper limit
${\cal A}_m$.

Classical and quantum arguments supporting the existence of a
maximal acceleration (MA) have long been discussed in the
literature \cite{prove}. MA also appears in the context of Weyl
space \cite{pap} and of a geometrical analogue of Vigier's
stochastic theory \cite{jv}. Its existence  would rid black hole
entropy of ultraviolet divergencies \cite{BHE},\cite{McG}, and
circumvent inconsistencies associated with the application of the
point-like concept to relativistic quantum particles \cite{he}.

Some authors regard ${\cal A}_m$ as a universal constant fixed by
Planck's mass \cite{b},\cite{infl}, but a direct application of
Heisenberg's uncertainty relations \cite{ca},\cite{pw} as well as
the geometrical interpretation of the quantum commutation
relations given by Caianiello, suggest that ${\cal A}_m$ be fixed
by the rest mass of the particle itself according to ${\cal
A}_m=2mc^3/\hbar$.

A limit on the acceleration also occurs in string theory. Here the
upper limit manifests itself through Jeans-like instabilities
\cite{gsv} which occur when the acceleration induced by the
background gravitational field is larger than a critical value
$a_c = (m\alpha)^{-1}$for which the string extremities become
causally disconnected \cite{gasp}. $m$ is the string mass and
$\alpha$ is the string tension. Frolov and Sanchez \cite{fs} have
then found that a universal critical acceleration $a_c =
(m\alpha)^{-1}$ must be a general property of strings.

While in all these instances the critical acceleration results
from the interplay of the Rindler horizon with the finite
extension of the particle \cite{emb},\cite{sa2}, in the Caianiello
model MA is a basic physical property of all massive particles
which appears from the out-set in the physical laws. At the same
time the model introduces an invariant interval in TM that leads
to a regularization of the field equations that does not require a
fundamental length as in \cite{qs} and does therefore preserve the
continuum structure of space-time.

Applications of Caianiello's model include cosmology \cite{infl},
where the initial singularity can be avoided while preserving
inflation, the dynamics of accelerated strings \cite{Feo}, the
energy spectrum of a uniformly accelerated particle \cite{emb},
the periodic structure as a function of momentum in neutrino
oscillations \cite{8} and the expansion of the very early universe
\cite{gasp}. The model also makes the metric observer--dependent,
as conjectured by Gibbons and Hawking \cite{Haw}.

The extreme large value that ${\cal A}_m$ takes for all known
particles makes a direct test of the model very difficult.
Nonetheless a direct test that uses photons in a cavity has also
been suggested \cite{15}. More recently, we have worked out the
consequences of the model for the classical electrodynamics of a
particle \cite{cla}, the mass of the Higgs boson \cite{Higgs} and
the Lamb shift in hydrogenic atoms \cite{lamb}. In the last
instance the agreement between experimental data and MA
corrections is very good for $H$ and $D$. For $He^+$ the agreement
between theory and experiment is improved by $50\%$ when MA
corrections are included. MA effects in muonic atoms also appear
to be measurable \cite{muo}.

In all these works space-time is endowed with a causal structure
obtained by means of an embedding procedure pioneered in \cite{8}
and discussed at length in \cite{emb},\cite{sch} . The procedure
stipulates that the line element experienced by an accelerating
particle, in the presence of gravity, is given by
 \begin{equation}\label{eq1}
 d\tau^2=\left(1+\frac{g_{\mu\nu}\ddot{x}^{\mu}\ddot{x}^{\nu}}{{\cal A}_m^2}
 \right)g_{\alpha\beta}dx^{\alpha}dx^{\beta}\equiv
 \sigma^2(x) g_{\alpha\beta}dx^{\alpha}dx^{\beta}\,{,}
 \end{equation}
where $\ddot{x}^{\mu}=d^2x^{\mu}/ds^2$ is the, in general,
non--covariant acceleration of a particle along its worldline.
 As a consequence, the effective space-time
geometry experienced by accelerated particles exhibits
mass-dependent corrections, which in general induce curvature,
give rise to a mass-dependent violation of the equivalence
principle and vanish in the classical limit $\left ({\cal
A}_m\right)^{-1} = {\hbar\over 2 m c^3}\rightarrow 0$.

We have recently studied the modifications produced by MA in the
motion of a test particle in a Schwarzschild field \cite{sch} and
found that these account for the presence of a spherical shell,
external to the Schwarzschild sphere, that is forbidden to any
classical particle and hampers the formation of a black hole. The
analogous occurrence of a classically impenetrable shell was
derived by Gasperini as a consequence of the breaking of the local
$SO(3,1)$ symmetry \cite{MG}. The shell remains impervious to
quantum, scalar particles \cite{qma}.

Before proceeding, a few comments are in order \cite{sch}. The
effective theory presented is intrinsically non-covariant, as is
the four-acceleration that appears in $\sigma^2(x)$. In addition
$\sigma^2(x)$ could be eliminated from (\ref{eq1}) by means of a
coordinate transformation if one applied the principles of general
relativity to this effective theory. On the contrary, the
embedding procedure requires that $\sigma^2(x)$ be present in
(\ref{eq1}) and that it be calculated in the same coordinates of
the unperturbed gravitational background. Nonetheless the choice
of $\ddot{x}^{\mu}$ is supported by the derivation of ${\cal A}_m$
from quantum mechanics, by special relativity and by the weak
field approximation to general relativity.  The model is not
intended, therefore, to supersede general relativity, but rather
to provide a way to calculate the quantum corrections to the
structure of space-time implied by (\ref{eq1}). Remarkably, the
results of \cite{sch} persist in isotropic coordinates \cite{qma}.

For convenience, the natural units $\hbar = c = G = 1$ are used
below.

 The purpose of this paper is to extend the calculation of the
MA corrections to the Reissner--Nordstr\"om metric. The problem is
not trivial. This metric, of form

\begin{equation}
ds^{2}=\left( 1-\frac{2M}{r}+\frac{e^{2}}{r^{2}}\right)
dt^{2}-\frac{dr^{2}}{1-\frac{2M}{r}+\frac{e^{2}}{r^{2}}}
-r^{2}\left(
d\vartheta ^{2}+\sin ^{2}\vartheta \, d\varphi ^{2}\right),
\label{Classical metric}
\end{equation}
does, in fact, contain two lengths, the mass of the central object
$M$ and its charge $e$ . The global space structure therefore
depends on the ratio of these two lengths. In addition, the
solution does not pertain to vacuum and contains a naked
singularity for certain values of the parameters.

Apart from the trivial case of vanishing $e$, which yields the
Schwarzschild metric, three cases must be distinguished. If
$\left| e\right| <M$, the Schwarzschild horizon moves from $2M$ to
$r_{A}=M+\sqrt{M^{2}-e^{2}}$, while a new horizon forms at the
origin. Its position is $r_{B}=M-\sqrt{M^{2}-e^{2}}$. In the
region between these two horizons, the component $g_{00}$ of the
metric tensor is negative and particles can only move towards the
origin. Elsewhere, particles are allowed to move away from the
origin.

When $\left| e\right| =M$, the two horizons merge at $r=M$. Yet
particles at this point cannot leave and the region inside the
sphere of radius $M$ is still not accessible to an external
observer.

Finally, for charges larger than the critical value, no horizon
exists, the metric (\ref{Classical metric}) can be used in the
whole space and particles can move in and out without
restrictions. The solution then corresponds to a naked
singularity.

In order to calculate the corrections to the Reissner--Nordstr\"om
field experienced by a  particle initially at infinity and falling
toward the origin along a geodesic, one must calculate the metric
induced by the embedding procedure discussed in \cite{sch} to
first order in the parameter ${{\cal A}^{-2}_m}$. On choosing
$\theta =\pi/2$, one finds
\begin{equation}
\sigma ^{2}\left( r\right) =1+\frac{1}{{\mathcal A}_m^{2}}\left[ \left( 1-\frac{%
2M}{r}+\frac{e^{2}}{r^{2}}\right) \ddot{t}^{ 2}-\left( 1-\frac{2M}{r}+\frac{%
e^{2}}{r^{2}}\right)
^{-1}\ddot{r}^{2}-r^{2}\ddot{\varphi}^{2}\right]. \label{Implicit
conformal factor}
\end{equation}

The components of the four--acceleration can be easily obtained by
following the same steps as in the Schwarzschild case \cite{sch}.
 When these are
substituted into (\ref{Implicit conformal factor}), the conformal
factor becomes $$ \sigma ^{2}\left( r \right)
=1+\frac{1}{{\mathcal A}_m^{2}}\left\{ -\frac {\left(
-\frac{M}{r^{2}}+\frac{e^{2}}{r^{3}}+\frac{\tilde{L}^{2}}{r^{3}}-\frac{3ML^{2}}{r^{4}}+\frac{2
e^{2}\tilde{L}^{2}}{r^{5}}\right)} {\left(
1-\frac{2M}{r}+\frac{e^{2}}{r^{2}}\right)}+ \right.  \\ $$
\begin{equation}
\left. +\left[ \tilde{E}^{2}\frac
{\left( \frac{2M}{r^{2}}-\frac{2e^{2}}{r^{3}}\right) ^{2}}
{\left( 1-\frac{2M}{r}+\frac{e^{2}}{r^{2}}\right) ^{3}}
-\frac{4\tilde{L}^{2}}{r^{4}}\right] \\
\left[\tilde{E}^{2}-\left( 1-\frac{2M}{r}+\frac{e^{2}}{r^{2}}\right)
\left( 1+\frac{\tilde{L}^{2}}{r^{2}}\right) \right] \right\}.\label{Explicit conformal factor}
\end{equation}
As in \cite{sch}, $\tilde{E}$ and $\tilde{L}$\ are the energy and
the angular momentum per unit of test particle mass.

Bearing in mind that the magnitude of the energy--momentum
four--vector is given by the rest mass $m$ of the particle
according to $g_{\alpha\beta}p^{\alpha}p^{\beta}=m^2$ and using
the modified metric (\ref{eq1}), one can introduce an effective
potential $V_{eff}$ by means of the equation
\begin{equation}
\left( \frac{dr}{d\tau}\right) ^{2}=\tilde{E}^{2}-V_{eff}^{2},
\end{equation}
where
\begin{equation}
V_{eff}^{2}=\tilde{E}^{2}-\frac{\tilde{E}^{2}}{\sigma ^{4}\left(
r\right) }+\frac{1}{\sigma ^{2}\left( r\right) }\left(
1-\frac{2M}{r}+\frac{e^{2}}{r^{2}}\right) \left(
1+\frac{\tilde{L}^{2}}{r^{2}\sigma ^{2}\left( r\right) }\right).
\label{Effective potential}
\end{equation}

In the limit ${\mathcal A}_m\rightarrow \infty $, or equivalently
$\sigma ^{2}\left( r\right) \rightarrow  1$, we recover the
classical Reissner--Nordstr\"om potential. Moreover, as
$e\rightarrow  0$, (\ref{Effective potential}) reduces to the
effective potential of \cite{sch}.

In the following we introduce the adimensional variable $\rho =
r/M$ and the parameters $\epsilon = \left(M {\cal
A}_m\right)^{-1}$, $\lambda = {\tilde L}/M$ and $\eta = e/M$.
 Notice that $V_{eff}^2(\rho)\rightarrow 1$ as $\rho \rightarrow
 \infty$ and $V_{eff}^2(\rho)\rightarrow \tilde{E}^{2}$ as
 $\rho \rightarrow 0$.
The main features of (\ref{Effective potential}) can  be derived
and compared with the corresponding corrections determined for the
Schwarzschild case.

a) $\left| \eta\right| <1$ and $\lambda = 0$. Since the component
$g_{00}=\left( 1-\frac{2}{\rho}+\frac{\eta^{2}}{\rho^{2}}\right) $
of the metric tensor vanishes at $\rho_{A} = 1 +
\sqrt{1-\eta^{2}}$ and $\rho_{B}=1 - \sqrt{1-\eta^{2}}$, it
follows from (\ref{Explicit conformal factor}) that the conformal
factor diverges as $ g_{00}^{-3}$ for the same values of $\rho$.
$\sigma ^{2}\left( \rho\right) $ always appears in the
denominators of the effective potential which thus equals
$\tilde{E}^{2}$.

Fig. 1 shows $V_{eff}^2\left(\rho\right)$ for a radially incoming
particle ($\lambda = 0$, $ \eta^{2} =0.1$) and different values of
the energy.

\begin{figure}[h]
\epsfysize= 9 cm
\begin{center}
\epsfbox{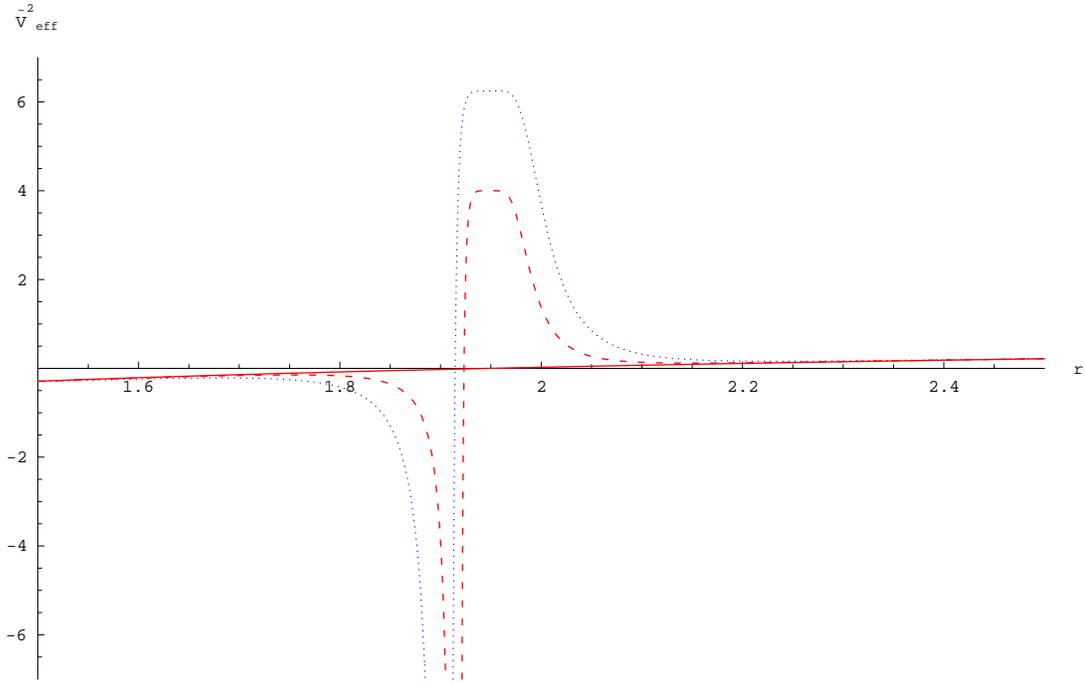} \caption{\footnotesize
{$V_{eff}^2\left(\rho\right)$ per unit of particle rest mass $m$
for ${\lambda} = 0$, ${\epsilon} = 10^{-3}$, $\eta^{2} = 0.1$,
$\tilde E = 2.5$ (dotted line) and $\tilde E = 2$ (dashed line).
The solid line refers to the usual Reissner Nordstr\"om
potential.}}
\end{center}
\end{figure}

\noindent As in the Schwarzschild case \cite{sch}, a shell is
formed in $\rho_{A}$ that prevents external particles from falling
into the no--return region.  The shell is dynamical in origin, is
impenetrable to classical particles and remains so at higher
orders of approximation in ${\cal A}_m^{-2}$.

 An expansion of (\ref{Effective
potential}) in the neighborhood of $\rho_{A}=1 + {\sqrt{1 - {{\eta
}^2}}}$ shows that the potential barrier behaves like
\begin{equation}
V_{eff}^2(\rho) =   {\tilde E^2} + {\frac{4\, f(\eta) }{{\tilde E^4}\,{{\epsilon }^2}}\,
      {{\left( \rho -\rho_{A}  \right)
      }^4}}  + O\left((\rho - \rho_{A})^{5}\right),
\end{equation}

where

$$
f(\eta) =
$$
$$\frac{\left(1-\eta^{2}\right) \, \left(1+{\sqrt{1 - {{\eta
}^2}}}\right)^{5}}{
     8 \left(32 - 72{{\eta }^2} + 54{{\eta }^4} - 15{{\eta
      }^6}\right) + 9{{\eta }^8} +{\sqrt{1 - {{\eta }^2}}}
 \left( 4 - {{\eta }^2} \right)
         \left( 64 - 96{{\eta }^2} + 36{{\eta }^4} - {{\eta }^6}
         \right)}.
         $$
$V_{eff}^2(\rho)$ clearly has the minimum $\tilde{E}^{2}$ on the
horizon
 $\rho_{A}$.
On the left of the barrier, there is a divergence originated by a
zero in $\sigma ^{2}\left( \rho\right) $. This divergence is
always negative because the dominant term in (\ref {Effective
potential}),
$
-\frac{\tilde{E}^{2}}{\sigma ^{4}\left( \rho\right) } $, is
negative for the corresponding values of $\rho$. To the left of
the divergence, the effective potential returns to values very
close to those of the Reissner-Nordstr\"om potential. The shape of
$V_{eff}^2$ is rather insensitive to variations of the parameters.
In particular, the barrier and the following divergence never
change.

Near $\rho_{B}$\ the situation is more complex since the conformal
factor has two, or four more roots (depending on the values of
$\eta$ and $\tilde E$), one on  the left of $\rho_{B}$ and one, or
three, on the right. The detailed behaviour of the corresponding
divergences is, however, irrelevant in the present context because
no classical particle can reach the region to the left of the
potential barrier near $\rho_{A}$.

b) $\left| \eta\right| <1$ and $\lambda \ne  0$. As in a) above,
the effective potential takes the value $\tilde E^{2}$ when $\rho$
tends to zero and to the horizons $\rho_{A}$ and $\rho_{B}$. On
the right of  $\rho_{A}$ the potential barrier is finite when the
angular momentum $\lambda$ is lower than a critical value
$\lambda_{c}(\epsilon, \tilde E, \eta)$, but infinite if $\lambda
> \lambda_{c}$. This barrier prevents external particles from
falling into the no--return region. Fig. 2 shows the behaviour of
$V_{eff}^2\left(\rho\right)$ in the neighborhood of the horizon
$\rho_{A}$ for two different values of $\lambda$,
$\lambda_{1}>\lambda_{c}$ and $\lambda_{2}<\lambda_{c}$.

\begin{figure}[h]
\epsfysize= 9 cm
\begin{center}
\epsfbox{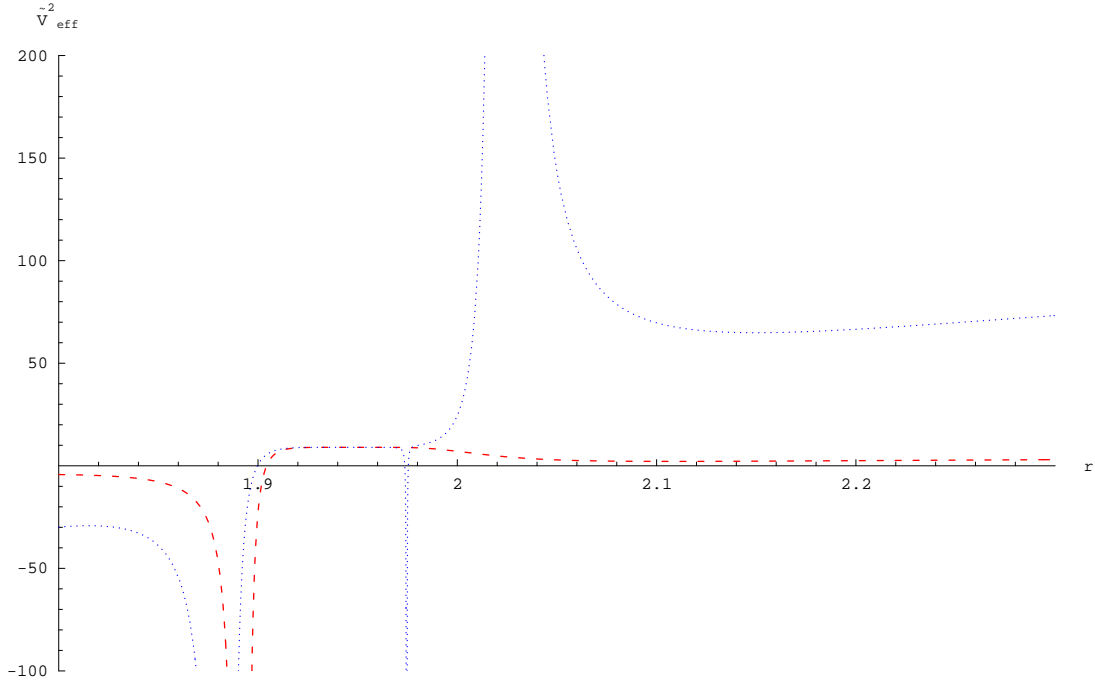} \caption{\footnotesize
{$V_{eff}^2(\rho)$ per unit of particle rest mass $m$ for $\tilde
E = 3$, ${\epsilon} = 10^{-3}$, $\eta^{2} = 0.1$, $\lambda_{1} =
50$ (dotted line) and $\lambda_{2} = 10$ (dashed line).}}
\end{center}
\end{figure}

The singularity structure on the left of $\rho_{A}$ is as complex
as in case a), and likewise irrelevant.

c) $\left| \eta\right| = 1$. At this critical value, $g_{00}$ has
a double root at $\rho=1$. Here, the effective potential forms the
usual barrier as $\sigma ^{2}\left(\rho\right)$ diverges. This is
shown in Fig. 3 (for $\lambda=0$) and in Fig. 4 (for $\lambda \ne
0$). An expansion of (\ref{Effective potential}) in the
neighborhood of $\rho_{A}=1$ yields the potential barrier $$
V_{eff}^2(\rho) =   {\tilde E^2} + {\frac{4}{{\tilde
E^4}\,{{\epsilon }^2}}\,
      {{\left( \rho -\rho_{A}  \right)
      }^6}}  + O\left((\rho - \rho_{A})^{7}\right)
$$
which, clearly, has the minimum $\tilde{E}^{2}$ on the  horizon
 $\rho_{A}=1$.
Near the origin there still is the singularity produced by the
zero of the conformal factor. Here too the sign is determined by
the values of $\tilde{E}$ and $\lambda$; it becomes positive when
$\lambda^{2} > \frac{{\tilde
E}^{2}\rho_{0}^{2}}{\left(1-\rho_{0}\right)^{2}}$, where
$\rho_{0}$ is the point at which $\sigma^{2}\left( \rho \right) $
vanishes.

\begin{figure}[h]
\epsfysize= 9 cm
\begin{center}
\epsfbox{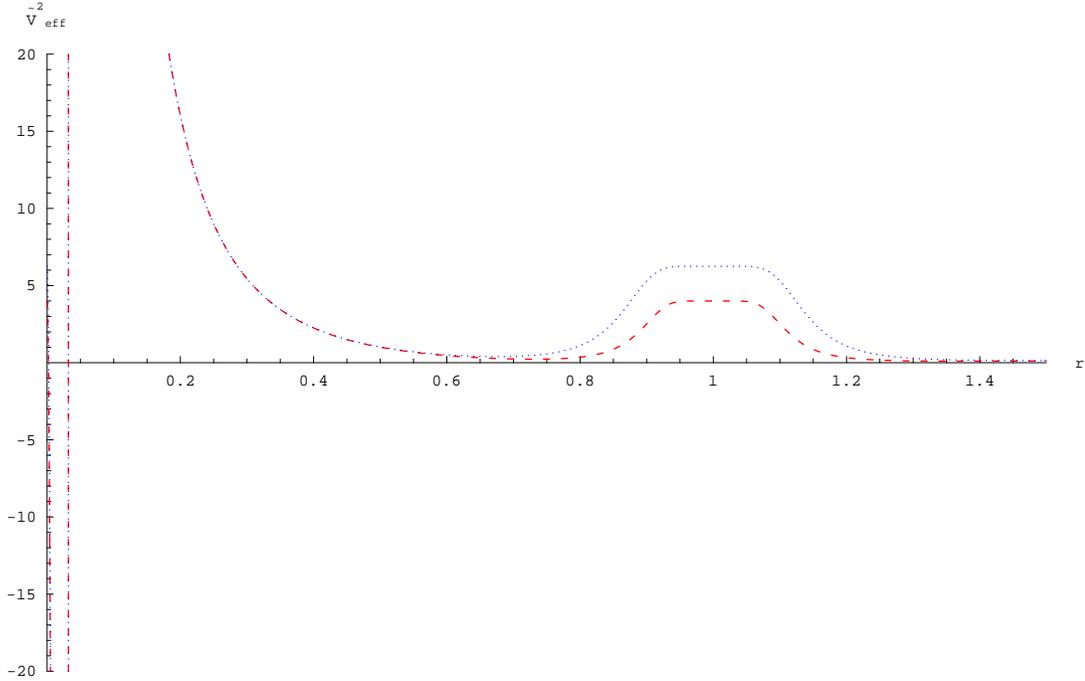} \caption{\footnotesize
{$V_{eff}^2(\rho)$ per unit of particle rest mass $m$ for $\lambda
= 0$, ${\epsilon} = 10^{-3}$, $\eta^{2} = 1$, $\tilde E = 2.5$
(dotted line) and $\tilde E = 2$ (dashed line).
 Note that
 there is only one singular point at $\rho\simeq 0.003 $, where
 $V_{eff}^2\rightarrow -\infty$}}
\end{center}
\end{figure}

\begin{figure}[h]
\epsfysize= 9 cm
\begin{center}
\epsfbox{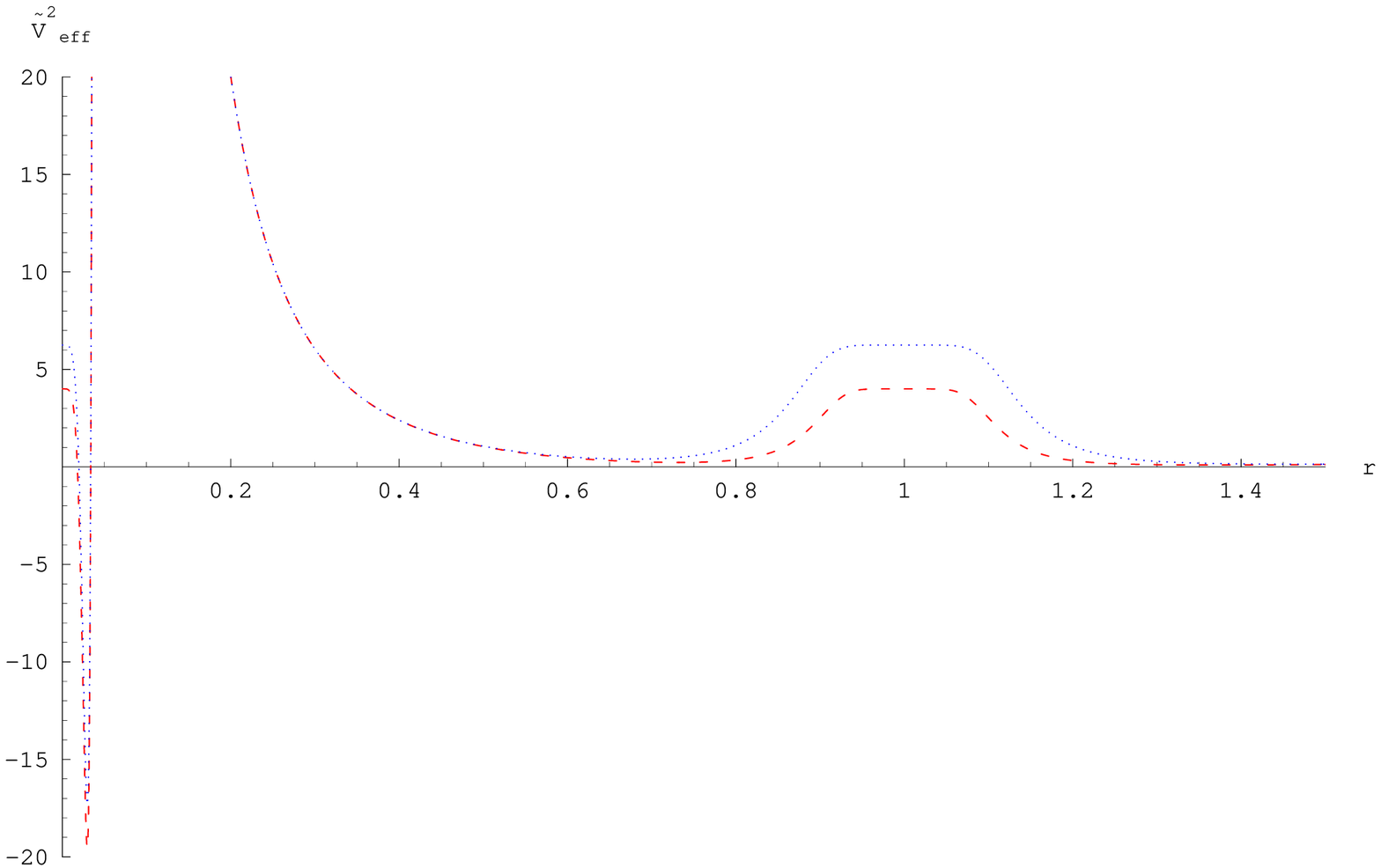} \caption{\footnotesize
{$V_{eff}^2(\rho)$ per unit of particle rest mass $m$ for $\lambda
= 0.1$, ${\epsilon} = 10^{-3}$, $\eta^{2} = 1$ and the values
$\tilde E = 2.5$ (dotted line) and $\tilde E = 2$ (dashed line).}}
\end{center}
\end{figure}
d) $\left|\eta\right| > 1$. $g_{00}$ is positive for all values of
$\rho$. There are no more horizons and the barrier shrinks (Fig.
5) until it disappears. Although the naked singularity is no
longer point-like, its behaviour remains unaffected by MA
corrections.
\begin{figure}[h]
\epsfysize= 9 cm
\begin{center}
\epsfbox{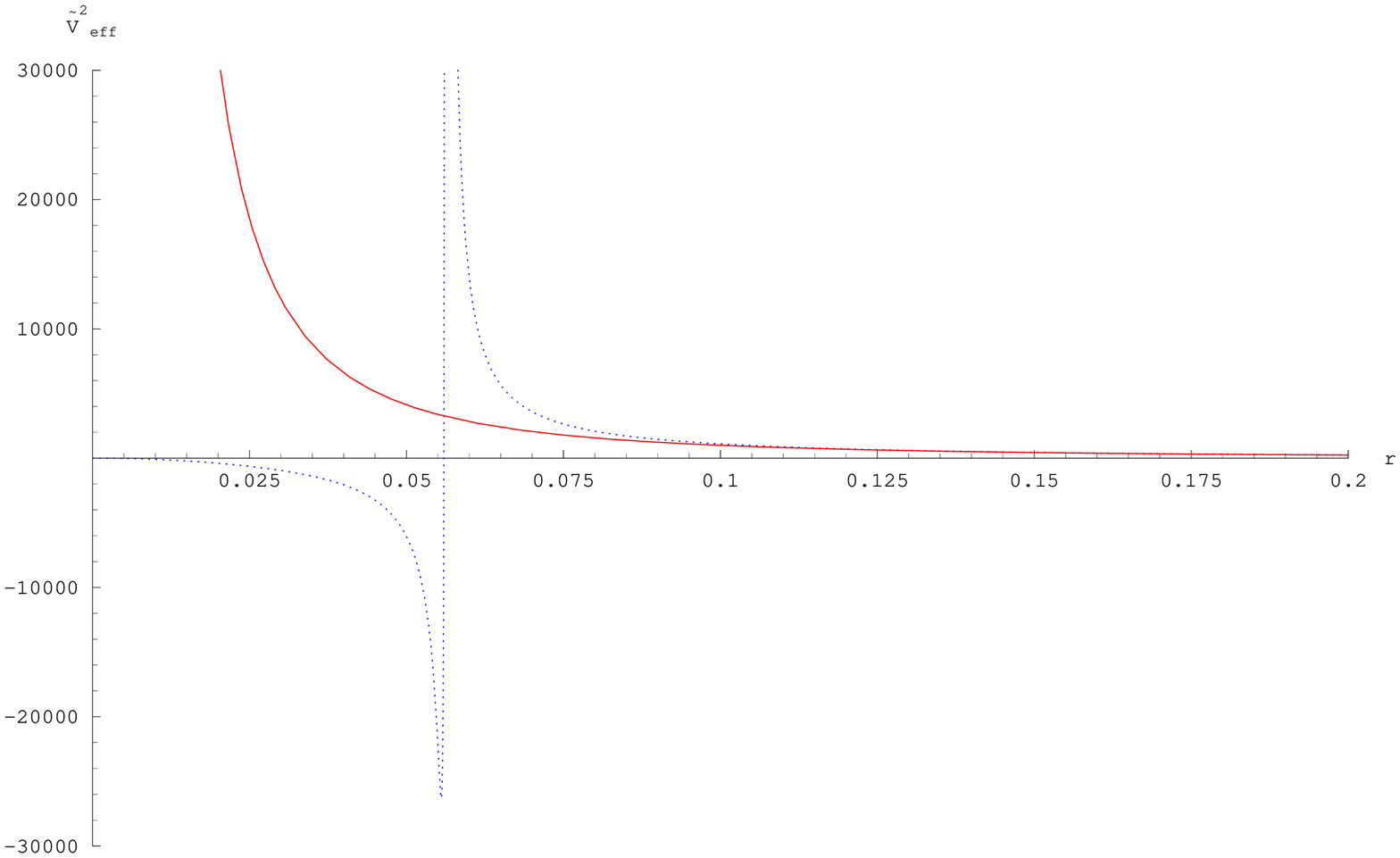} \caption{\footnotesize
{$V_{eff}^2(\rho)$ per unit of particle rest mass $m$ for $\lambda
= 10^{-2}$, ${\epsilon} = 10^{-3}$, $\eta^{2} = 10$ and $\tilde E
= 2.5$ (dotted line). The solid line represents the Reissner
Nordstr\"om potential for the same values of the parameters.
 }}
\end{center}
\end{figure}

For a black hole with 3.3 solar masses, the critical value for the
central charge is $5.7\times 10^{20}$ coulomb. Obviously, this
value can never be reached in the normal case irrespective of the
relative signs of the charges of $m$ and $M$. If, in fact, matter
and black hole had charges of the same sign, then the
electromagnetic repulsion would prevent the fall of matter. If the
signs were opposite, then particles would be strongly attracted
and would tend to neutralize the whole system. For this reason
real black holes can only be ``low charged'' and should be
described by the $\left| e\right| < M$ case.

The corrections to the Reissner--Nordstr\"om metric induced by the
presence of MA do indeed represent an interesting confirmation of
the physics already discussed for the Schwarzschild case. Apart
from the classical shift from $\rho=2$ to $\rho=\rho_{A}$, the
possible presence of charge in the source essentially leaves the
barrier at the external horizon unchanged. This means that all the
remarks about the formation of black holes made in \cite{sch} with
regard to the Schwarzschild metric can be extended to the ``low
charged'' Reissner--Nordstr\"om space. Classically, the presence
of the barrier forbids the formation of a black hole by the
accretion of matter, unless the latter is transformed first into
massless particles and these are absorbed by the star at a rate
higher than the corresponding re-emission rate. A recent
calculation \cite{qma} shows that even the quantum properties of
matter do not alter this behaviour and that gravitational collapse
would at least be slowed down. Nonetheless the shell would still
retain the appearances of a very intensely radiating, charged,
compact object ($\rho_A \leq 2$).

Besides confirming what found in \cite{sch} about black hole
dynamics, the application of Caianiello model to the
Reissner--Nordstr\"om metric offers other interesting aspects.

Since the shell is impervious to charged matter of both signs, the
charge of the source can not change. It is not possible, at
present, to speculate on the behaviour of electron-positron pairs
produced in the immediate vicinity of a source with sufficiently
high charge.

A barrier on a horizon is always accompanied by a singularity on
the side where $g_{00}$\ is negative. These divergences are even
present in the curvature invariants, so they must be considered as
physical singularities of the effective metric. They are however
inaccessible to massive probes and only regard the behaviour of
matter inside a black hole, if the latter somehow formed. It is
however doubtful that the Reissner--Nordstr\"om solution would
adequately describe this type of matter.

 For high values of the angular momentum the divergence
in the effective potential becomes positive causing the formation
of an infinite, repulsive potential barrier.

It is important to remember that the Reissner--Nordstr\"om
space--time allows causality violations in cases c) (and d)) and
in the region $\rho < \rho_B$ of a) and b)\cite{Car}.
 The impenetrability
of the shell renders inaccessible those regions with physically
inadmissible properties that violate causality.

 Finally, MA alters
the radius of the naked singularity, but not its nature and does
not therefore embody any form of cosmic censorship.

\bigskip
\bigskip

\begin{centerline}
{\bf Acknowledgments}
\end{centerline}

Research supported  by Ministero dell'Universit\`a e della Ricerca
Scientifica of Italy, funds ex 40\% and 60\% DPR 382/80 and the
Natural Sciences and Engineering Research Council of Canada.

\end{document}